\documentclass[manuscript]{aastex63}
\usepackage{graphicx}	
\usepackage{amsmath}	
\usepackage{amssymb}	
\usepackage{subfigure}
\usepackage{float}
\usepackage{ae,aecompl,amsmath,amsbsy,mathtools, nccmath}
\usepackage{epstopdf}
\usepackage{multirow}
\usepackage{hyperref}
\usepackage{lipsum}
\usepackage{multirow}
\usepackage{booktabs}

\graphicspath{{./}{figures/}}
\begin{document}

\title{Modelling ion acceleration and transport in corotating interaction regions: the mass-to-charge ratio dependence of the particle spectrum}

\correspondingauthor{Gang Li}
\email{gangli.uah@gmail.com}

\author[0000-0002-9829-3811]{Zheyi Ding}
\affiliation{School of Geophysics and Information Technology, China University of Geosciences (Beijing), Beijing 100083, China}
\affiliation{Centre for mathematical Plasma Astrophysics, KU Leuven, 3001 Leuven, Belgium}

\author[0000-0003-4695-8866]{Gang Li}
\affiliation{Department of Space Science and CSPAR, University of Alabama in Huntsville, Huntsville, AL 35805, USA}

\author[0000-0001-6344-6956]{Nicolas Wijsen}
\affiliation{Centre for mathematical Plasma Astrophysics, KU Leuven, 3001 Leuven, Belgium}

\author[0000-0002-1743-0651]{Stefaan Poedts}
\affiliation{Centre for mathematical Plasma Astrophysics, KU Leuven, 3001 Leuven, Belgium}

\author[0000-0003-4267-0486]{Shuo Yao}
\affiliation{School of Geophysics and Information Technology, China University of Geosciences (Beijing), Beijing 100083, China}

\begin{abstract}
We investigate the role of perpendicular diffusion in shaping energetic ion spectrum 
in corotating interaction regions (CIRs), focusing on its mass-to-charge ($A/Q$) dependence. We simulate a synthetic CIR using the EUropean Heliospheric FORcasting Information Asset (EUHFORIA) and model the subsequent ion acceleration and transport by solving the focused transport equation incorporating both parallel and perpendicular diffusion. Our results reveal distinct differences in ion spectra between scenarios with and without perpendicular diffusion. In the absence of perpendicular diffusion, ion spectra near CIRs show a strong $(A/Q)^{\epsilon}$ dependence with $\epsilon$ depending on the turbulence spectral index, agreeing with theoretical predictions.  In contrast, the incorporation of perpendicular diffusion, characterized by a weak $A/Q$ dependence, leading to similar spectra for different ion species. This qualitatively agrees with observations of energetic particles in CIRs.
\end{abstract}

\section{Introduction}
\label{sec:intro}

Corotating interaction regions (CIRs) form when the faster solar wind overtakes slower solar wind. The stream interaction leads to compression waves and eventually the formation of the forward-reverse shock pair. These persistent, large-scale structures are often associated with tens of keV/n to several MeV/n energetic particles in the inner heliosphere \citep{Richardson2018LRSP...15....1R}. Early work by \citet{Fisk1980ApJ...237..620F} assumed that CIRs accelerate particles beyond several au, where shocks are generally formed. Later it was argued that compression regions associated with CIRs can accelerate ions in a similar way to shocks, but without the need of shocks \citep{Giacalone2002ApJ...573..845G}.
While Ulysses observations have seen energetic ions at the reverse shock as far as $3$ au \citep{Mason1999SSRv...89..327M},  recent observations at 1 au have pointed to a local source for CIR-related energetic ions, specifically those below approximately 1 MeV/n \citep{Mason2008ApJ...678.1458M,Ebert2012ApJ...749...73E,wijsen2021}. 
These findings suggest that the energy spectrum observed by L1 spacecraft results from a combination of local acceleration processes and transport effects as particles move inward from distant shocks. Previous work by \citet{Mason2008ApJ...678.1458M} has examined the abundance ratios of heavy ions versus their energy/nucleon, revealing a near constancy below $2\;$MeV/n. This constancy is also mirrored in spectral rollover energies ($E_{0}$). A more recent study by \citet{Filwett2019ApJ...876...88F} found that the rollover energy in stream interaction regions (SIRs) does not show a significant mass-to-charge ($A/Q$) ratio dependence, a stark contrast to solar energetic particle (SEP) events, where a varying $A/Q$ dependence of ion time profiles and/or spectra are usually seen \citep{mason2006ApJ...647L..65M,Desai2016ions}. 

Given that the acceleration and the subsequent transport of energetic ions from CIRs is diffusive in nature, the presence or absence of the $A/Q$ dependence in ion rollover energies can be related to the $A/Q$ dependence  of the ion diffusion coefficients.
In considering the transport of ions in the solar wind, the ion mean free path ($\lambda$) can be expressed as a power law in rigidity ($R=A/Qv$), $\lambda \sim R^{\alpha}$, implying a diffusion coefficient ($\kappa=\frac{1}{3} v \lambda $) that scales as $(A/Q)^{\alpha}v^{(\alpha+1)}$ \citep{Cohen2003AdSpR,Ellison1985ApJ...298..400E}. Different ions of different energies will have similar time profiles if their diffusion coefficients are the same. This is often referred to as the equal diffusion coefficient condition \citep[e.g.,][]{mason2006ApJ...647L..65M,Mason2012ApJ...761..104M}.  In considering the acceleration process, if $\kappa$ has a certain $A/Q$ dependence, it will translate to a rollover energy that depends on $A/Q$ \citep{Cohen2005JGRA..110.9S16C}. \citet{Li2009ApJ...702..998L} suggested that the shock geometry plays a critical role in determining the $A/Q$ dependence in gradual SEP events, with the total diffusion coefficient as a function of shock obliquity.
In the work of \citet{Li2009ApJ...702..998L}, the rollover energy of shock-accelerated ions is shown to 
have a  $(A/Q)^\epsilon$ dependence with $\epsilon=-2$  for parallel shocks and $\epsilon=-1/5$ for perpendicular shocks. Later observations of large SEP events associated with interplanetary shocks \citep{Desai2016ions,Desai2016ApJ..Fe} found the spreading of $\epsilon$ in good agreement with  \citet{Li2009ApJ...702..998L}.
If the observer is not at the acceleration site, transport may complicate the $A/Q$ dependence of the rollover energy. \citet{Zhao2016ApJ...821...62Z} studied how particle transport with pitch angle scattering might contribute to the $A/Q$ dependence observed in heavy ion spectra during SEP events.  
These studies have underscored the significance of diffusion coefficient and the shock geometry in understanding the $A/Q$ dependence of the ion acceleration and transport.  

Unlike SEP events studied in \citet{Desai2016ions,Desai2016ApJ..Fe}, where the shock geometry varies as a shock propagates out and where the transport effects also vary with time as  the distance between the shock and the observer decreases, CIRs are mostly stationary structures in large scales.  Therefore, the roles of ion acceleration and transport in shaping the CIR energetic particle spectrum are less entangled and are easier to understand than SEP events. However, at small scales (several to tens of ion inertial lengths), shock front can be irregular and non-stationary. Recent observations and kinetic simulations by \citet{Kajdic2019ApJL,Guo2021FrASS,Trotta2023MNRAS,Trotta2023ApJL} have shown that small-scale irregularities are ever-present at shock front, evolving in space and time. These structures can be crucial in understanding the injection of superthermal particles because the injection depends on the shock geometry \citep{Li+2012}, and as shown in \cite{Trotta2023ApJL}, the local shock obliquity shows high variability along a irregular shock front, leading to an inhomogeneous injection of the superthermal particles. These results suggest that the shock properties in the large-scale magnetohydrodynamics (MHD) simulations, where small scale irregularities are ignored, can only be regarded as some mean properties of the shock.

Assuming the interplanetary magnetic field (IMF) takes the form of nominal Parker field, and ignoring perpendicular diffusion and small-scale shock irregularities, then any given observer will connect to a single point at the CIR shock. Under such an assumption, \citet{Zhao2016JGRA..121...77Z} examined how energetic CIR particle spectra depend on observer's longitude. This picture, however, needs to be revised if we consider the effect of perpendicular diffusion. Particle perpendicular diffusion has two contributions. The first is the meandering of the field line due to the presence of the magnetic fluctuation $\delta \mathbf{B}$, which contains no $A/Q$ dependence. This has been examined by, e.g., \citet{Bian+2021, BianLi2022,Li2023ApJ...945..150L}.  Because the field
line itself is ``diffusive", when particles move along these field lines, there will be displacements perpendicular to the unperturbed background field. Therefore, perpendicular diffusion is also referred to cross-field diffusion (with respect to the unperturbed field). Under the zero-gyroradius approximation, particle perpendicular diffusion is the same as that of the meandering field line \citep{Bian+2021, BianLi2022}. 
The second contribution is the finite gyroradius effect. Retaining a non-zero gyroradius,  one can evaluate particle's trajectory and compute its perpendicular displacement from the original unperturbed field lines. 
\citet{Matthaeus2003ApJ...590L..53M} developed a Non-Linear-Guiding-Center (NLGC) theory to compute particle's perpendicular diffusion coefficient where the underlying turbulence is assumed to be of ``slab+2D" geometry.
This work was later extended to other turbulence geometries \citep{shalchi2010analytic, shalchi+2009+book}.

Previous simulations of CIR events 
\citep{Zhao2016JGRA..121...77Z,wijsen2021} have not explicitly and systematically considered the effect 
of perpendicular diffusion. This motivates our current study. 
In this letter, we investigate how perpendicular diffusion can influence ion acceleration and transport in a CIR event, and potentially lead to a weak 
$A/Q$ dependence of ion spectra. 
In Section~\ref{sec:model}, we provide a brief discussion of our modelling approach. We first describe the procedure of simulating a synthetic CIR using the EUropean Heliospheric FORcasting Information Asset (EUHFORIA; \citet{pomoell2018}) model. We then describe how the particle acceleration and transport is followed from the vicinity of the CIR reverse shock to an observer. This follows closely to recent works by \citet{wijsen2021,Wijsen2023JGRA..12831203W}. In section~\ref{sec:results},  we present a comparative analysis for cases with and without perpendicular diffusion. This comparison focuses on the ion fluxes and spectra at 1 au. 
Conclusion is given in Section~\ref{sec:conclusion}.

\section{MODEL}
\label{sec:model}
A CIR is modelled by using EUHFORIA, which solves the ideal MHD equations from 0.1 au to 5 au. Following \citet{wijsen2019a}, a steady CIR structure is achieved by setting a uniform solar wind speed of $400\;$km/s at the inner boundary, with the exception of a specific region where the solar wind speed is elevated to $750\;$km/s. This region is defined by the condition $(\phi - 270^{\circ})^2 + (\theta - 0^{\circ})^2 < (20^{\circ})^2$, where $\phi$ and $\theta$ are the longitude and latitude, respectively. For simplicity, we assume a monopolar positive magnetic field at the inner boundary to avoid heliospheric current sheet. With this configuration, 
 the interaction between the slow and fast streams leads to the formation of a CIR, characterized by forward and reverse shock waves.

To examine the acceleration and transport of energetic particles, 
 we follow a similar approach as \citet{wijsen2021} and inject seed ions of $50\;$keV/n with a radial dependence of $r^{-2}$ to the reverse shock. This, of course, is a simplification, as we 
 remarked earlier that the injection energy and efficiency of seed ions depend on the local shock geometry and small-scale shock irregularities \citep{Trotta2023ApJL} can affect the shock geometry. We then follow these particles by solving the three-dimensional focused transport equation through the EUHFORIA solar wind domain in the corotating frame \citep{wijsen2019a}, expressed as:
\begin{equation}\label{eq:fte}
\begin{array}{r@{}l}
\frac{\partial f}{\partial t}+(\mathbf{V}_{\rm sw}+\mu v\mathbf{b})\cdot \nabla f + \frac{1-\mu^2}{2}\left [ v\nabla \cdot \mathbf{b}  +\mu \nabla \cdot \mathbf{V}_{\rm sw} - 3\mu \mathbf{b} \mathbf{b} : \nabla \mathbf{V}_{\rm sw}\right ]\frac{\partial f}{\partial \mu}\\
+ p\left[\frac{1-3\mu^2}{2}(\mathbf{b} \mathbf{b} : \nabla \mathbf{V}_{\rm sw}) - \frac{1-\mu^2}{2}\nabla\cdot \mathbf{V}_{\rm sw} \right ]\frac{\partial f}{\partial p}\\
= \frac{\partial}{\partial \mu}\left (D_{\mu \mu} \right )\frac{\partial f}{\partial \mu}+ \nabla \cdot (\boldsymbol{\kappa}_{\perp} \cdot \nabla f)  . %

\end{array}
\end{equation}
In this equation, the gyro-averaged particle distribution function is represented by $f(\mathbf{x}, p,\mu, t)$, which is a function of  spatial coordinate $\mathbf{x}$, momentum magnitude $p$, the pitch-angle cosine $\mu$ and time $t$. Additional elements in equation~\eqref{eq:fte} include the particle velocity $v$, the velocity of the solar wind $\mathbf{V}_{\rm sw}$, and the unit vector of the magnetic field $\mathbf{b}$. Both $\mathbf{V}_{\rm sw}$ and $\mathbf{b}$ are taken from a single steady-state snapshot from the EUHFORIA model, so $\mathbf{V}_{\rm sw}$ and $\mathbf{b}$ are time-independent in the corotating frame.
The two terms on the right-hand side of equation~\eqref{eq:fte} describe particle diffusion. These are the pitch-angle diffusion given by $D_{\mu\mu}$ (related to the parallel diffusion coefficient $\kappa_{\parallel}$) and the perpendicular diffusion, represented by the diffusion tensor $\boldsymbol{\kappa}_\perp$. These terms collectively model the influence of a fluctuating magnetic field, $\delta \boldsymbol{B}$, on particle transport.  Additionally, we adopt the assumption by \citet{Wijsen2023JGRA..12831203W} for the radial dependence of $\delta B^2/B^2$. 

Following  \citet{ding2022A&A...668A..71D}  in describing  $D_{\mu \mu}$ using the  the quasi-linear theory (QLT)  \citep{Jokipii+1966} and $\kappa_{\perp}$ from $\kappa_{\parallel}$ using the NLGC Theory \citep{shalchi2010analytic}, we can obtain the dependence of $A/Q$ in $\kappa_{\parallel}$ and $\kappa_{\perp}$ in the form of $\kappa_{\parallel} \sim (A/Q)^{\epsilon_{\parallel}}$ and $\kappa_{\perp} \sim  (A/Q)^{\epsilon_{\perp}}$. Here $\epsilon_{\parallel}$ is decided by the spectral index of the turbulence power spectral index $\beta$ ($I(k) \sim k^{\beta}$).  Using QLT and NLGC theory, \citet{Li2009ApJ...702..998L} showed that $\epsilon_{\parallel} = \beta+2$ and $\epsilon_{\perp} = (\beta+2)/3$ . Clearly, $\kappa_{\perp}$ has a weaker $A/Q$ dependence than $\kappa_{\parallel}$. A recent statistical study by \citet{Park2023ApJL} found that $\beta$ ranges from $-1.4$ to $-2.0$, both downstream and upstream of the fast forward and fast reverse interplanetary shock, with a mean value around $-1.7$. 
In this work, we assume $\beta$ to be $-5/3$ when considering the transport of energetic ions in the solar wind, away from the reverse shock.  This corresponds to a Kolmogorov spectrum. Close to the shock, besides the case of $\beta = -5/3$, 
we also consider a control case of $\beta = -2$. In this case, one finds both $\kappa_{\parallel}$ and $\kappa_{\perp}$ are independent of ion species.  The turbulence level near a CIR and its reverse/forward shock can be enhanced 
\citep{Crooker1999cir..book..179C,Richardson2004SSRv..111..267R}. Indeed,  wave intensity across interplanetary shocks are often enhanced. A case study of the large October 29 2003 event \citep{Li2005AIPC..781..233L} showed an enhancement of more than a factor of $10$.  Recent work by \citet{Park2023ApJL} suggested that the turbulence level across fast shocks can increase, on average, by $3-5$ times.
From the rarefaction region, the spectral index $\beta$ does not differ much from 
the Kolmogorov value. This is because the CIR shocks are quasi-perpendicular in geometry, therefore there is not much amplification of waves due to streaming ions, as in the case of parallel shocks \citep{Li+2003,Li+2005ions}.

The simulation duration is taken to be $27$ days, approximately one solar rotation.  At the inner and outer boundary, a reflective and an absorbing boundary condition are used, respectively. 
The steady-state solution at a desired position is obtained for a continuous time-independent injection by convoluting the Green's function solution at different times.

\section{Results}
\label{sec:results}
\begin{figure}[ht!]
	\epsscale{1.1}
	\plotone{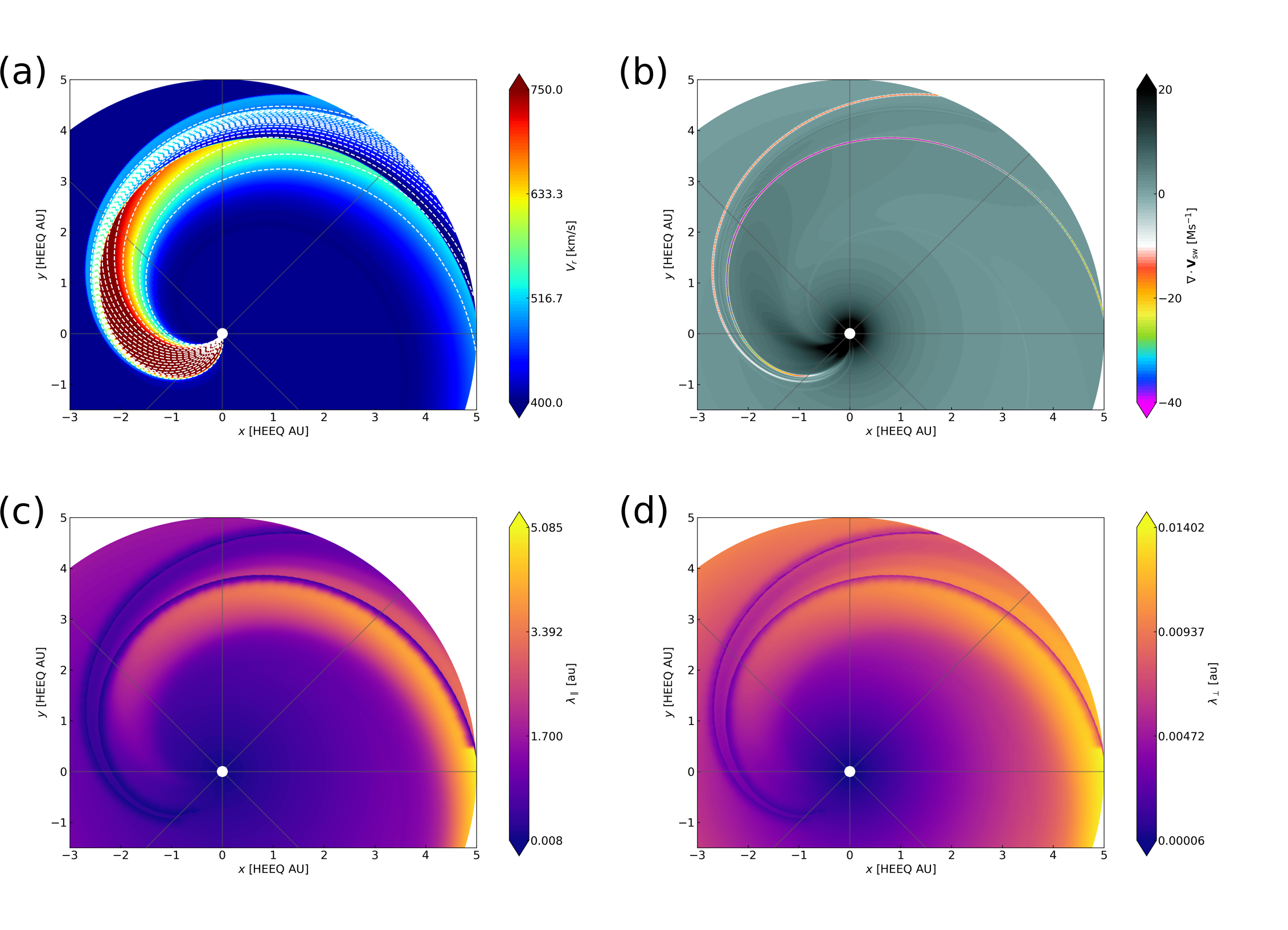}
	\caption{The solar wind radial speed and the assumed parallel and perpendicular mean free paths for $50$ keV proton in the solar equatorial plane. Panel (a) shows the radial solar wind speed. White dashed lines are magnetic field lines from the simulation. Panel (b) shows the divergence of the solar wind velocity. The compression waves can be recognized as the colorful spiral-shaped structures. Panels (c) and (d) show, respectively, the parallel and perpendicular mean free paths for $50\;$keV proton. The simulation domain is from $r=0.1$ au to $r=5$ au.}  \label{Fig1}
\end{figure}

 All analyses are done in the equatorial plane, as in the recent work by \citet{Wijsen2023JGRA..12831203W}. Figure~\ref{Fig1}(a) displays the radial velocity of the solar wind, $V_r$. 
Slow and fast winds can be seen clearly in the plot.
Figure~\ref{Fig1}(b) plots $\nabla \cdot \mathbf{V}_{\rm sw}$. 
 Since the solar wind propagates radially out, $\nabla \cdot \mathbf{V}_{\rm sw}$ is positive if $V_{\rm sw}$ is a constant, but can become negative near stream interface. In Figure~\ref{Fig1}(b), these negative divergence regions include both the forward and  reverse shocks.  The gray shaded regions indicate expanding solar wind areas.
  In this study,  we limit the injection only at the reverse shock as this allows for a clean analysis of the particle spectra.
We note that the reverse shock is more efficient in accelerating particles \citep{Lee1982SSRv...32..205L}. 
Figure~\ref{Fig1}(c) shows the parallel mean free path $\lambda_{\parallel}$ for $50\;$keV proton. Away from the reverse shock, 
$\lambda_{\parallel}$ increases with the heliocentric distance, reflecting the gyro-frequency of particles and their wave-particle resonance conditions with the background magnetic field \citep{Wijsen2023JGRA..12831203W}. Note that in the rarefaction region particles exhibit a larger parallel mean free path. Figure~\ref{Fig1}(d) shows the resulting perpendicular mean free path $\lambda_{\perp}$ for $50\;$keV proton, computed from $\lambda_{\parallel}$ (see details in \citet{ding2022A&A...668A..71D}). The similar azimuthal variation of $\lambda_{\perp}$ is due to its dependence on $\lambda_{\parallel}^{1/3}$. Near the reverse shock, the mean free path decreases. This is because near the compression regions the wave activity increases due to stream interactions. We note that EUHFORIA simulations do not capture how the upstream wave intensity evolves. To model the transport of energetic particles, a prescription of the wave intensity, which amounts to a description of particle diffusion coefficient, is needed \citep{wijsen2021,Wijsen2023JGRA..12831203W}.
Recently, in examining two energetic storm particle events observed by Solar Orbiter,  \citet{ding2024modelling} assumed that the wave intensity upstream of the shock decays exponentially from the shock. In this work, we follow a similar approach to describe the wave intensity upstream of the CIR shocks. By way of example, we approximate the wave intensity $I(k,\mathbf{x})$ upstream of the reverse shock by,
\begin{equation}\label{eq:db2_shock}
I\left(k_{},\mathbf{x}\right)\,=\,I_{0}\left(k_{},\mathbf{x}_{}\right)+aI_{0}\left(k_{},\mathbf{x}_{0}\right)\cdot \exp\left(-\frac{|\mathbf{x}-\mathbf{x}_{0}|}{L}\right),
\end{equation}
where $k$ is the wave number; $\mathbf{x}$ is the position upstream of the shock and $\mathbf{x}_0$ is the closest point to $\mathbf{x}$ on the shock; 
$I_0$ is the ambient wave intensity and $a$ is the amplitude of enhanced wave intensity at the shock, taken to be $5$ in our simulation; $L$, the diffusion length scale, is assumed to be $0.05\;$au. Note that $L$ could be energy-dependent and radial-dependent, similar to the case considered at CME-driven shocks \citep{Li+2005ions,Li2021,ding2022modellingEastWest,ding2024modelling}. 
Upstream of the shock, $I(k)$ reduces exponentially to the ambient level in a distance of $\sim L$.  Downstream of the shock, where $\nabla \cdot \mathbf{V}_{\rm sw} <0$, we assume $I\left(k_{},\mathbf{x}\right)\,=\,(a+1)I_{0}\left(k_{},\mathbf{x}_{0}\right)$, which is a constant.

Since we follow the evolution of the particle distribution function with the focused transport equation, acceleration is implicitly taken care of when particles cross the shock. In fact, as shown in \citet{wijsen2019a,wijsen2021,Wijsen2023JGRA..12831203W}, shock is not necessary
for accelerating particles, as only a negative divergence is needed \citep{Giacalone2002ApJ...573..845G}. 
Our simulation follows closely to \citet{wijsen2019a,wijsen2021,Wijsen2023JGRA..12831203W}, but with the consideration of the additional turbulence near the compression region, as shown in equation~(\ref{eq:db2_shock}).  We remark that the ion acceleration efficiency and their maximum energies positively correlate with the turbulence level near the CIR. It could be important to understand some large CIR events with maximum energy up to $\sim$ 20 MeV/n \citep{Mason2008ApJ...678.1458M,Richardson2004SSRv..111..267R}.

\begin{figure}[ht!]
	\epsscale{1.1}
	\plotone{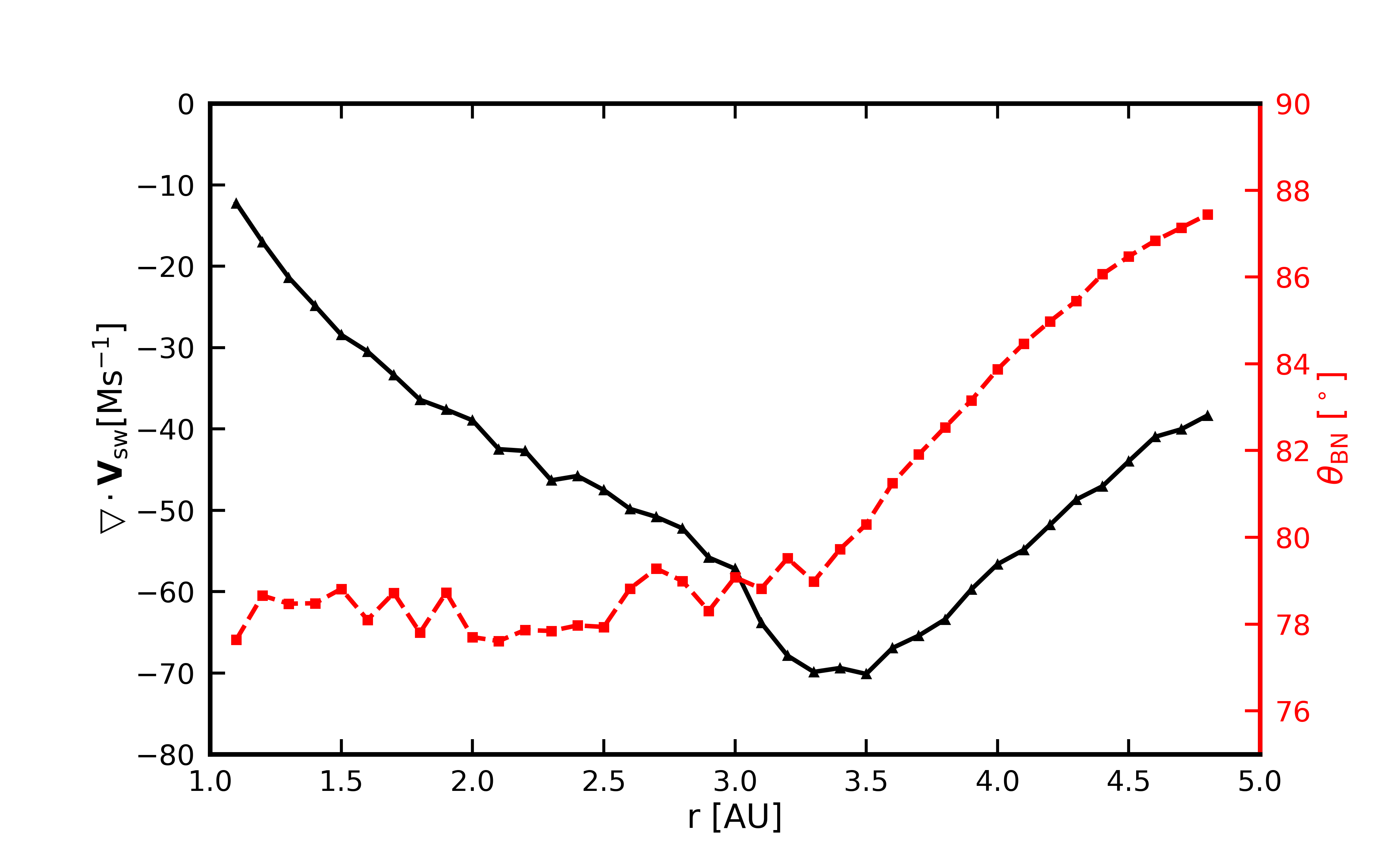}
	\caption{The divergence of the solar wind velocity $\nabla \cdot \mathbf{V}_{\rm sw}$ and the shock obliquity $\theta_{\rm BN}$  as a function of radial distance along the CIR reverse shock.}  \label{Fig2}
\end{figure}

Figure~\ref{Fig2} shows the divergence of the solar wind velocity, $\nabla \cdot \mathbf{V}_{\rm sw}$, and the shock obliquity, $\theta_{\rm BN}$, as functions of radial distance along the reverse shock. Notably, the magnitude of $\nabla \cdot \mathbf{V}_{\rm sw}$ initially increases with radial distance, reaching a maximum around 3.5 au, before gradually decreases. This implies a radial variation in the strength of the reverse shock, suggesting a more efficient particle acceleration occurring around 3.5 au. This phenomenon could potentially correlate with the observed radial dependence of ion intensity during CIR events, where a positive correlation between ion intensities and radial distances has been recently noted \citep{Allen2021GeoRL..4891376A}. A more comprehensive exploration of this relationship will be pursed in a future study. The value of $\theta_{\rm BN}$ shows a gradual increase from $78^{\circ}$ to $88^{\circ}$ as one moves from 1 au to 5 au, indicative of the shock geometry becoming more perpendicular further out. The obliquity of the CIR shock is crucial for understanding ion acceleration. As discussed by \citet{Li2009ApJ...702..998L}, the total diffusion coefficient $\kappa$ can be generally  expressed as,
\begin{equation}\label{eq:total_kappa}
\kappa(v) = \kappa_{\parallel,0}(v/v_0)^{\gamma_{\parallel}}\left( \frac{A}{Q} \right)^{\epsilon_{\parallel}}\cos^2(\theta_{\rm BN}) + \kappa_{\perp,0}(v/v_0)^{\gamma_{\perp}}\left( \frac{A}{Q} \right)^{ \epsilon_{\perp}}\sin^2(\theta_{\rm BN}),
\end{equation}
where $\kappa_{\parallel,0}$ and $\kappa_{\perp,0}$ represent the parallel and perpendicular diffusion coefficients for protons at speed $v_0$, $\gamma_{\parallel}$ and $\gamma_{\perp}$ are the indices denoting the dependence on proton speed and $\epsilon_{\parallel}$ and $\epsilon_{\perp}$ refer to the $A/Q$ dependence. In the QLT and NLGC theory, one can obtain  $\gamma_{\parallel}=(\beta+3)$ 
 and $\gamma_{\perp}=(\beta+5)/3$ \citep{ding2022A&A...668A..71D}. As discussed in Section~\ref{sec:model}, $\epsilon_{\parallel}= \beta+2$ and $\epsilon_{\perp}=(\beta+2)/3$. Following the equal diffusion coefficient conditions \citep{Mason2012ApJ...761..104M},   the $A/Q$ dependence of rollover energy $E_0$ is $E_{0} \sim (A/Q)^{-2(\beta+2)/(\beta+3)}$ at a parallel shock and is $E_{0} \sim (A/Q)^{-2(\beta+2)/(\beta+5)}$ at a perpendicular shock. This dependence at an oblique shock is determined by the interplay between $\kappa_{\perp,0}(v)/\kappa_{\parallel,0}(v)$ and $\theta_{\rm BN}$ (see details in \citet{Li2009ApJ...702..998L}).  Consequently, for $\beta=-5/3$, it translates to $\epsilon_{\parallel}=1/3$ and $\epsilon_{\perp}=1/9$, a significant $(A/Q)^{-1/2}$ dependence is expected predominantly in parallel shocks. However, given that the reverse shock primarily exhibits a quasi-perpendicular geometry, a weak $A/Q$ dependence is anticipated.

\begin{figure}[ht!]
	\epsscale{1.1}
	\plotone{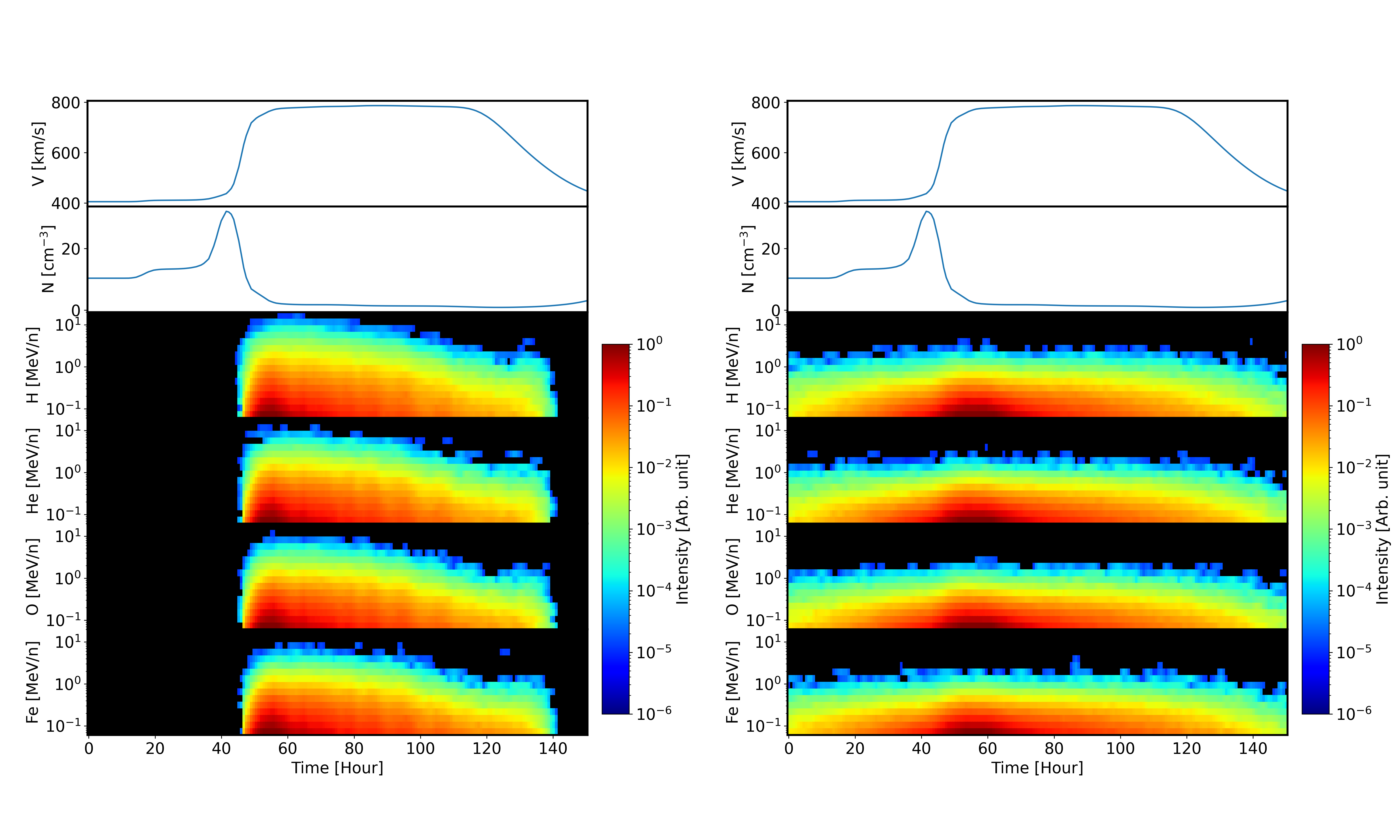}
	\caption{The left and right panels show from top to bottom: solar wind speed, solar wind number density, the spectrogram of normalized energetic ion flux intensities versus energy and time (proton, helium, oxygen, iron) for an observer at 1 au. 
 Left (right) panels are for the case without (with) perpendicular diffusion. }  \label{Fig3}
\end{figure}

To understand the impact of perpendicular diffusion on the $A/Q$ dependence during CIR events, we compare simulation results at 1 au with and without the inclusion of perpendicular diffusion. Figure~\ref{Fig3} presents the time profiles of solar wind speed and number density, and spectrograms of normalized ion flux intensity as functions of energy and time, including proton (H), helium (He) , oxygen (O), and iron (Fe).  The values of $A/Q$ for H, He, O, and Fe are chosen to be $1/1$, $4/2$, $16/6$ and $56/12$, respectively \citep{Desai2016ions}. To better compare the spectrogram and energy spectra, the injection of different ions is assumed to be the same. The observer is fixed in the inertial frame.  
The left panels of Figure~\ref{Fig3} display simulation results for the case without perpendicular diffusion. In this case, a clear peak in ion intensity is observed near the compression region, followed by a gradual decay. The decay phase reflects the fact that the observer connects to further and further parts of the shock (the CIR structure is rotating counter-clock wise in the inertial frame) where the injection drops. Note that ions accelerated beyond 1 au need to propagate back to 1 au. Because no perpendicular diffusion is included, a fixed observer in the corotating frame only connects to a single point at the shock. A fixed observer in the inertial frame, of course, connects to different parts of the shock at different times, yet the connection point moves along the shock with a constant rotation. This is vastly different from the case when we include the perpendicular diffusion, which  is shown in  
the right panels of Figure~\ref{Fig3}. 
We see now the peaks near the compression regions are smeared out. Energetic ions appear $\sim 40$ hours prior to the compression, and over a total of $120$ hours (30 hours before and 90 hours after the compression), the intensities maintain relatively plateau-like shape.  
This suggests that the inclusion of the perpendicular diffusion allows an observer to ``probe" a large portion of the shock surface at one time instead of only ``probe" a single point on the shock surface at a time.

\begin{figure}[ht!]
	\epsscale{1.1}
	\plotone{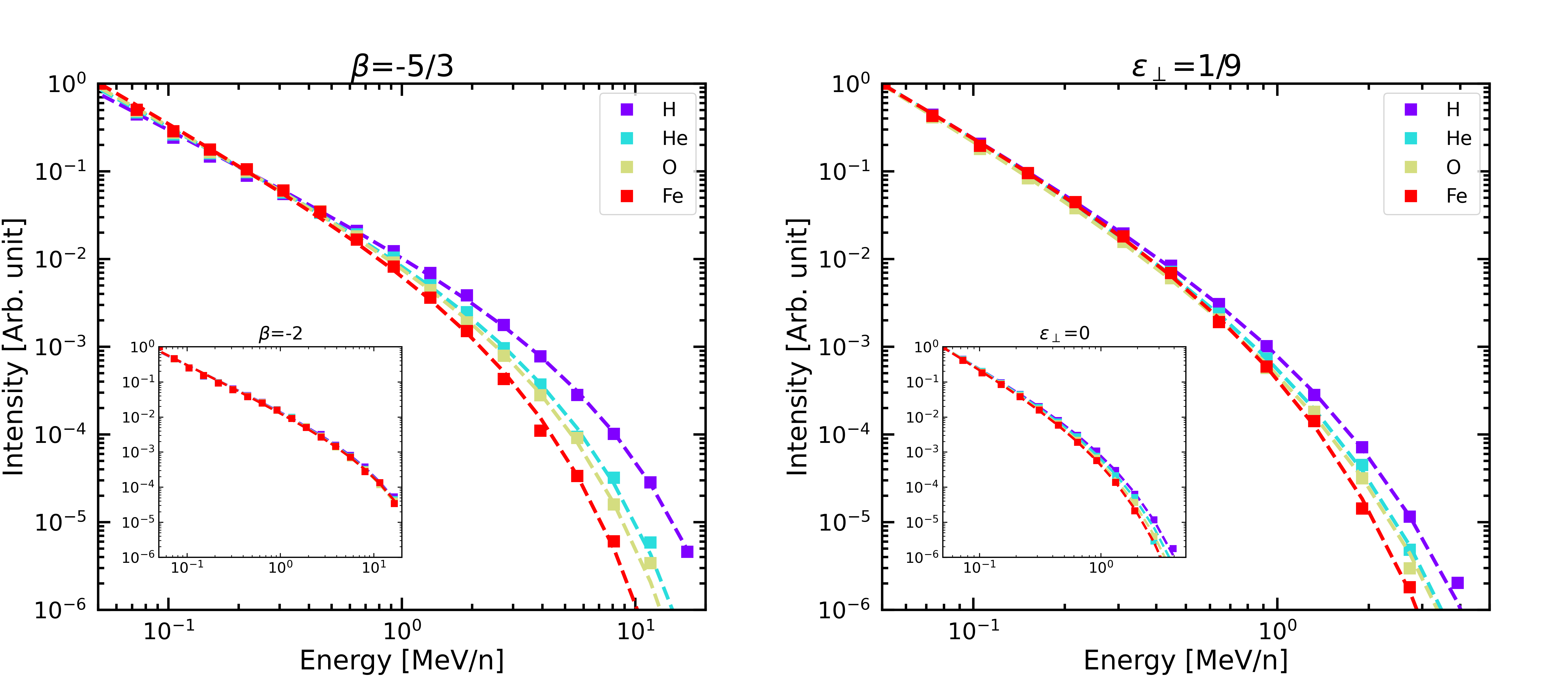}
	\caption{The normalized time-integrated ion spectra for an observer at 1 au. The left and right panels represent the case without and with perpendicular diffusion , respectively.  
 In the left main panel $\beta=-5/3$, and in the inset $\beta=-2$;
 in the right main panel $\epsilon_{\perp}=1/9$,  and in the inset $\epsilon_{\perp}=0$.
 The dash lines show the fitted spectra. See text for details.}  \label{Fig4}
\end{figure}

Figure~\ref{Fig4} compares the time-integrated ion spectra near the compression region: from $45\;$hours to $60\;$hours.  This period is chosen such that the  observed particles are predominantly accelerated near 1 au, reducing the transport effect in the solar wind. In the left panels, where  perpendicular diffusion is not included, the ion spectra exhibit good power-law-like behavior with an exponential decay. The spectra toward the higher energy end show clear $A/Q$ dependence which is also reflected in the fitted rollover energy. We fit each ion spectra using a form of a single power law with an exponential tail, $E^{-\gamma}\exp (-E/E_0)$, where $\gamma$ is the spectral index and $E_0$ is the rollover energy.   Dash lines represent the fitting results.
In this case, the rollover energies for different ions are $E_0^{\rm H}=4.06 \pm 0.16\;$MeV/n, $E_0^{\rm He}=2.84 \pm 0.12\;$MeV/n, $E_0^{\rm O}=2.54 \pm 0.09\;$MeV/n, and $E_0^{\rm Fe}=1.79 \pm 0.10\;$MeV/n, which shows $E_0 \sim (A/Q)^{-0.51 \pm 0.04}$.  
Since no perpendicular diffusion is included,  this $A/Q$ dependence is governed solely by $\kappa_{\parallel}$ and it is directly linked to the turbulence spectral index $\beta$, 
taken to be $-5/3$ here. The theoretical prediction of rollover energy  with only $\kappa_{\parallel}$ is  $E_{0} \sim (A/Q)^{-2(\beta+2)/(\beta+3)}$ indicated by equation~\eqref{eq:total_kappa}.   Our fitting result is consistent with the prediction, $E_{0} \sim (A/Q)^{-1/2}$ when $\beta=-5/3$.  In the case of $\beta=-2$, the theory predicts no $A/Q$ dependence. This is shown in the inset of left panel. We do not show the corresponding spectrogram of ion intensity in Figure~\ref{Fig3} for $\beta=-2$ since these are less informative.  For $\beta >-5/3$, the $A/Q$ dependence becomes more significant (not shown here). 

The right panels in Figure~\ref{Fig4} show the spectra with perpendicular diffusion included.
In the main panel, where $\beta=-5/3$,  so using the NLGC theory, we have  $\kappa_{\perp} \sim (A/Q)^{1/9}$. Following equation~\eqref{eq:total_kappa}, this leads to a $E_{0} \sim (A/Q)^{-1/5}$ if only perpendicular diffusion dominates the acceleration process.
Different ion spectra in the main panel show only minor differences above $1\;$MeV/n comparing to the left main panel. The fitted rollover energies shows a $E_{0} \sim (A/Q)^{-0.27\pm 0.07}$.  This is slightly 
stronger than $(A/Q)^{-1/5}$, since $\kappa_{||}$ is also playing a role. 
We also examine the case where 
$\kappa_{\perp}$ is  $A/Q$-independent (i.e., $\epsilon_{\perp}=0$). Such a choice corresponds to 
considering only the field line meandering contribution to $\kappa_{\perp}$. The resulting spectra 
are shown in the inset. As expected, it leads to an even less pronounced $A/Q$ dependence, characterized by $(A/Q)^{-0.18\pm 0.02}$. Again, this $A/Q$ dependence is attributed to $\kappa_{\parallel}$. As $\beta \rightarrow -2$,  then $A/Q$ dependence diminishes, as indicated in the left sub-panel.

Figure~\ref{Fig4} is the most important result of this work. It demonstrates clearly the crucial role of perpendicular diffusion in regulating the accelerated ion spectrum near CIR shocks. It is interesting to note that the maximum energy of ions when perpendicular diffusion is included  is lower than that when perpendicular diffusion is ignored. This is because perpendicular diffusion can effectively move particles away from the shock region and reduce the acceleration efficiency. This is similar to the behavior of the seed population when perpendicular diffusion is included \citep{Wijsen2023JGRA..12831203W}.

\section{Conclusion}
\label{sec:conclusion}

In this study, we have explored the influence of perpendicular diffusion on ion acceleration and transport in CIRs, particularly focusing on the implications for the $A/Q$ dependence in ion rollover energies. In the absence of perpendicular diffusion, ion spectra near the compression region exhibit a pronounced $A/Q$ dependence in rollover energies, which is related to the turbulence power spectral index. However, the inclusion of perpendicular diffusion largely removes  such a $A/Q$ dependence, leading to a markedly different behavior. The spectra of different ion species do not differ much and the rollover energies of ions are similar. Our findings highlight the crucial role of perpendicular diffusion in the ion acceleration near CIR shocks. It implies that CIR environment could be the best place to examine perpendicular diffusion.  Observations of CIR energetic particle spectrum  \citep{Mason2008ApJ...678.1458M,Filwett2019ApJ...876...88F} showed no strong $A/Q$ dependence. 
We suggest that this is because of a weak $A/Q$ dependence of $\kappa_{\perp}$, and the fact that CIR shocks are largely quasi-perpendicular. Our results provide a model basis for interpreting the $A/Q$ dependence of ion spectra in CIR observations. 

\acknowledgments
The work at UAH is supported by NSF ANSWERS grant 2149771 and NSF SHINE2301365. GL also acknowledges supports from ISSI and ISSI-BJ through the international team 581. N.W.\ acknowledges support from the Research Foundation - Flanders (FWO-Vlaanderen, fellowship no.\ 1184319N). SP acknowledges support from the projects C14/19/089 (C1 project Internal Funds KU Leuven), G.0B58.23N and G.0025.23N (WEAVE) (FWO-Vlaanderen), 4000134474 (SIDC Data Exploitation, ESA Prodex-12), and Belspo project B2/191/P1/SWiM.
SY acknowledges support from National Natural Science Foundation of China under contract No.\ 42074204 and the 111 Plan Overseas Expertise Introduction Project for Discipline Innovation under contract No. B20011. For the computations we used the infrastructure of the VSC–Flemish Supercomputer Center, funded by the Hercules foundation and the Flemish Government–department EWI.

\bibliographystyle{aasjournal}
\bibliography{ccc}{}

\end{document}